\documentclass[aps,pre,a4paper,twocolumn]{revtex4-2}
\pdfoutput=1

\usepackage{psfrag}
\usepackage{amssymb,amsmath,amsthm}
\usepackage[dvips]{graphicx}
\usepackage{verbatim}
\usepackage{color}
\usepackage{bm}

\usepackage{xcolor}
\definecolor{red}{rgb}{0,0,0}
\colorlet{r}{red}
\definecolor{blue}{rgb}{0,0,0}
\colorlet{b}{blue}
\newcommand{\rc}[1]{\textcolor{b}{#1}}

\begin{document}

\begin{widetext}
\begin{center}{\textit{Published in Phys.~Rev.~E \textbf{106}, 054609 (2022). {\copyright}2022 American Physical Society.}}
\end{center}
\end{widetext}

\title{The role of boundaries for displacements and motion in two-dimensional fluid or elastic films and membranes}

\date{\today}

\author{Tyler Lutz}

\email{tyler.lutz@ovgu.de}    
\affiliation{Institut f\"ur Physik, Otto-von-Guericke-Universit\"at Magdeburg, Universit\"atsplatz 2, 39106 Magdeburg, Germany}

\author{Sonja K.\ Richter}
    
\affiliation{Institut f\"ur Physik, Otto-von-Guericke-Universit\"at Magdeburg, Universit\"atsplatz 2, 39106 Magdeburg, Germany}

\author{Andreas M.\ Menzel}
    
\email{a.menzel@ovgu.de}
\affiliation{Institut f\"ur Physik, Otto-von-Guericke-Universit\"at Magdeburg, Universit\"atsplatz 2, 39106 Magdeburg, Germany}

\begin{abstract}
Thin fluid or elastic films and membranes are found in nature and technology, for instance, as confinements of living cells or in loudspeakers. 
When applying a net force, resulting flows in an unbounded two-dimensional incompressible low-Reynolds-number fluid or displacements in a two-dimensional linearly elastic solid seem to diverge logarithmically with the distance from the force center, which has led to some debate. Recently, we have demonstrated that such divergences cancel when the total (net) force vanishes. Here, we illustrate that, if a net force is present, the boundaries play a prominent role. Already a single no-slip boundary regulates the flow and displacement fields and leads to their decay to leading order inversely in distance from a force center and the boundary. In other words, it is the boundary that stabilizes the system in this situation, unlike the three-dimensional case, where an unbounded medium by itself is able to absorb a net force. We quantify the mobility and displaceability of an inclusion as a function of the distance from the boundary, as well as interactions between different inclusions. In the case of free-slip boundary conditions, a kinked boundary is necessary to achieve stabilization.
\end{abstract}

\maketitle


\section{Introduction} \label{intro}

In the past, several studies addressed the scenario of imposing locally concentrated forces on the bulk of incompressible fluids under low-Reynolds-number conditions \cite{kim1991microhydrodynamics, dhont1996introduction} and linearly elastic solids \cite{landau1986theory}. Mainly, this is achieved by inserting discrete objects into continuous media. External forces are exerted on these objects that transmit these forces to their surroundings. If other immersed or embedded discrete objects are present, they are exposed to the flow or deformation fields. Corresponding interactions between the inclusions as mediated by their environment result. Overall, such situations were investigated for viscous fluids \cite{mazur1982many, kim1991microhydrodynamics, dhont1996introduction, hoell2019multi}, elastic solids \cite{phan1994load, kim1995faxen, schopphoven2019elastic, puljiz2017forces, puljiz2019displacement}, and to some degree for viscoelastic media \cite{schmiedeberg2005one, puljiz2019memory, richter2021rotating}. 

In three-dimensional bulk systems, the flows or displacements introduced by a localized force center decay inversely with the distance from that force center. This spatial dependence is reflected by the associated Green's function for the underlying equations, which vanishes at infinite distance \cite{kim1991microhydrodynamics, dhont1996introduction, kachanov2003handbook}. Remarkably, the situation is different for two-dimensional systems. Effectively, the equations of motion for thin fluid films can be reduced to two-dimensional inplane spatial variations \cite{oron1997long}. Moreover, such situations arise when the equations of linear elasticity for thin elastic membranes are in effect reduced to two dimensions \cite{landau1986theory}\rc{, specifically when assuming so-called plane stresses}. Apart from such systems of small thickness, also the description of three-dimensional systems of vanishing spatial derivatives along the third dimension reduces to two dimensions \rc{in states of so-called plane strain}.  

In contrast to the three-dimensional case, a two-dimensional system is not stable against a net force acting on its inside. Specifically, the flows or displacements diverge logarithmically with the distance from a concentrated force center \cite{squires2006breaking}, which is also reflected by the associated Green's function \cite{phan1983image}. Previously, it was argued that this signals a breakdown of the linear theory and that nonlinear contributions then become important \cite{proudman1957expansions}. In a previous work, we have taken a different point of view and demonstrated that the logarithmic divergence does not appear, if the net forces on the system sum up to zero \cite{richter2022mediated}. An example are inclusions that pairwise exert forces onto each other according to Newton's third law, which they then transmit to the fluid or elastic two-dimensional environment, implying vanishing net force. 

Here, we address what actually happens in two-dimensional situations when a net force does act at a certain position onto the two-dimensional system. In fact, it turns out that the boundaries play a central role in this case. Naturally, any real system has a boundary and needs to be fixed or clamped in some way, if overall displacement is to be avoided. The diverging Green's function in two dimensions indicates that the fixed boundary becomes crucial, no matter how far away the force center is located. At the same time, we show that one fixed no-slip boundary in an otherwise infinitely extended two-dimensional system is sufficient to stabilize the situation and to cancel the divergence. In other words, there is no divergence in the two-dimensional half-space under no-slip boundary conditions. Fluid flows or displacements vanish at infinite distance from the boundary and the force center. Still, the underlying equations remain completely linear and nonlinearities are not necessary to regulate the system. Similar regularization is achieved for free-slip boundary conditions, if a kink is introduced at the boundary. 

We begin by overviewing the underlying equations for the continuous medium in Sec.~\ref{sec_green}, where we also introduce the appropriate Green's function in the presence of one no-slip boundary for an otherwise infinitely extended two-dimensional system. Next, we interpret in Sec.~\ref{sec_displ} the result concerning the mobility and displaceability  of one inclusion within the system as a function of the distance from the boundary and the direction of displacement. In Sec.~\ref{sec_mediated}, we address the mutual interactions between inclusions mediated through the fluid or elastic environment. We add related remarks concerning free-slip boundary conditions at a kinked boundary in Sec.~\ref{sec_freeslip}. 
In Sec.~\ref{sec_free-slip_mediated}, we evaluate mediated interactions in the case of the free-slip geometry. Finally, we conclude in Sec.~\ref{sec_concl}.

\section{Green's function for the two-dimensional half-space}
\label{sec_green}

We present in the following the solution by analyzing the underlying equations for displacements in an isotropic, homogeneous, linearly elastic medium. The solution for flows in an incompressible fluid under low-Reynolds-number conditions is formally obtained by setting in these results the Poisson ratio, associated with the compressibility of linearly elastic materials, to $\nu\rightarrow 1/2$. Moreover, the shear modulus $\mu$ in the elasticity solution needs to be replaced by the viscosity $\eta$ and the displacement field $\mathbf{u}(\mathbf{x})$ by the velocity field of the fluid flow $\mathbf{v}(\mathbf{x})$ in transferring the results to low-Reynolds-number hydrodynamics. In these expressions, $\mathbf{x}$ refers to the spatial position as measured from the origin of our coordinate system. 

We start from the Navier--Cauchy equations
\begin{equation}\label{eq_NC}
\nabla^2\mathbf{u}(\mathbf{x})+\frac{1}{1-2\nu}
\bm{\nabla}\bm{\nabla}\cdot\mathbf{u}(\mathbf{x})
={}-\frac{1}{\mu}\mathbf{f}(\mathbf{x}),
\end{equation}
so that $\mathbf{f}(\mathbf{x})$ represents the force density acting on the inside of the elastic material. \rc{In the following, we interpret these equations strictly in two dimensions, so that $\mathbf{x}$ marks the position vector in the Cartesian $xz$-plane. Likewise, both  $\mathbf{u}(\mathbf{x})$ and $\mathbf{f}(\mathbf{x})$ are considered as two-dimensional in-plane vector fields.} The associated Green's function follows by solving Eq.~(\ref{eq_NC}) for $\mathbf{f}(\mathbf{x})=\mathbf{F}\,\delta(\mathbf{x}-\mathbf{x}_0)$, where $\delta$ denotes the Dirac delta function and $\mathbf{F}$ is a force vector. 

\begin{figure}
 		\centering 
\includegraphics[width=8.5cm]{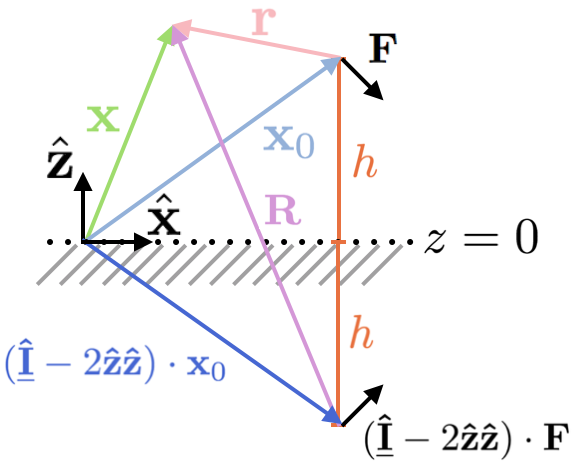}
 		\caption{\rc{Illustration of the geometry considered for a no-slip boundary at $z=0$. A force $\mathbf{F}$ acts at a point-like force center at position $\mathbf{x}$ and distance $h$ from the boundary on a semi-infinite elastic medium that occupies the half-space $z\geq0$. We obtain the image position of the mirrored force $(\underline{\mathbf{\hat{I}}}-2\mathbf{\hat{z}}\mathbf{\hat{z}})\cdot\mathbf{F}$ via $(\underline{\mathbf{\hat{I}}}-2\mathbf{\hat{z}}\mathbf{\hat{z}})\cdot\mathbf{x}_0$. Besides, we introduce the abbreviations $\mathbf{r}$ and $\mathbf{R}$ as defined in Eqs.~(\ref{eq_rxx0}) and (\ref{eq_R}), respectively.}}
 		\label{Fig:scheme}
 	\end{figure}

The elastic system fills the half-space $z\geq0$. A no-slip boundary is introduced at $z=0$, that is, 
\begin{equation}
\mathbf{u}(z=0)=\mathbf{0}.   
\end{equation}
The point force $\mathbf{F}$ is positioned at a distance $h$ from the boundary, which implies $\mathbf{\hat{z}}\cdot\mathbf{x}_0=z_0=h$. \rc{ Figure~\ref{Fig:scheme} illustrates the geometry we envision.} 
Additionally, we set 
\begin{equation}\label{eq_rxx0}
\mathbf{r}=\mathbf{x}-\mathbf{x}_0,
\end{equation}
so that $\mathbf{r}$ actually measures the spatial position relative to the position of the point force. We introduce the image location that results from mirroring the position $\mathbf{x}_0$ at $z=0$. To this end, we employ the reflection operator $\underline{\mathbf{\hat{I}}}- \rc{2}\mathbf{\hat{z}}\mathbf{\hat{z}}$, where $\underline{\mathbf{\hat{I}}}$ denotes the unit matrix. Thus, the position of the image beyond the boundary at $z=0$ becomes $(\underline{\mathbf{\hat{I}}}-\rc{2}\mathbf{\hat{z}}\mathbf{\hat{z}})\cdot\mathbf{x}_0=\mathbf{x}_0-2h\mathbf{\hat{z}}$. By further definition, 
\begin{equation}\label{eq_R}
\mathbf{R}=\mathbf{r}+2h\mathbf{\hat{z}}
=\mathbf{x}-(\underline{\mathbf{\hat{I}}}-\rc{2}\mathbf{\hat{z}}\mathbf{\hat{z}})\cdot\mathbf{x}_0
=\mathbf{x}-\mathbf{x}_0+2h\mathbf{\hat{z}}
\end{equation}
measures spatial positions relative to the location of the image. 

To derive the Green's function $\mathbf{\underline{B}}(\mathbf{r})$, we follow the scheme outlined for three-dimensional systems \cite{menzel2017force}. We split the Green's function as
\begin{equation}\label{eq_ansatzB}
\mathbf{\underline{B}}(\mathbf{r}) =
\mathbf{\underline{G}}(\mathbf{r})
-\mathbf{\underline{G}}(\mathbf{R})
+\mathbf{\underline{W}}(\mathbf{R}), 
\end{equation}
where $\mathbf{\underline{G}}(\mathbf{r})$ denotes the Green's function for the infinitely extended two-dimensional system
\begin{equation}\label{eq_G}
		\mathbf{\underline{G}}(\mathbf{r})=  \frac{1}{8\pi(1-\nu)\mu}\left[-(3-4\nu)\ln (r)\, \underline{\hat{\mathbf{I}}}+\frac{\mathbf{r}\mathbf{r}}{r^2}\right] .
\end{equation}
$\mathbf{r}\mathbf{r}$ marks the dyadic product and $r=|\mathbf{r}|$. 
The displacement field follows as $\mathbf{u}(\mathbf{x})
=\mathbf{\underline{B}}(\mathbf{x}-\mathbf{x}_0)\cdot\mathbf{F}$. 

Equation~(\ref{eq_ansatzB}) \rc{first} includes through $\mathbf{\underline{G}}(\mathbf{r})$ the consequences of \rc{the physically present} point-like force center \rc{acting at position} $\mathbf{x}_0$ \rc{on the elastic material}. \rc{Next,} through $-\mathbf{\underline{G}}(\mathbf{R})$, \rc{we include the mirrored} point-like counter-force center\rc{. That is, we mirror the position $\mathbf{x}_0$ at $z=0$ to $\mathbf{x}_0-2h\mathbf{\hat{z}}$ and there locate a counter-force} of equal magnitude but opposite orientation.  \rc{Yet, the displacements induced by the physical and the image force do not satisfy the no-slip boundary conditions at $z=0$. We therefore introduce via $\mathbf{\underline{W}}(\mathbf{R})$ an additional} image system at $\mathbf{x}_0-2h\mathbf{\hat{z}}$ to ensure that the no-slip boundary condition at $z=0$ is met. $\mathbf{\underline{W}}(\mathbf{R})$ must for $z\geq0$ satisfy Eq.~(\ref{eq_NC}) for $\mathbf{f}(\mathbf{x})=\mathbf{0}$. We obtain $\mathbf{\underline{W}}(\mathbf{R})$ by Fourier-transforming Eq.~(\ref{eq_NC}) as well as the boundary condition with respect to the transverse coordinate \cite{blake1971note}. On this way, we benefit from introducing an auxiliary field $\mathbf{P}(\mathbf{R})=\bm{\nabla}\cdot\mathbf{\underline{W}}(\mathbf{R})$ \cite{menzel2017force}. In Fourier space, we can solve for the searched-for quantities and determine remaining coefficients from the boundary conditions \cite{menzel2017force}. The final result is obtained after inverse transformation as
\begin{eqnarray}
\mathbf{\underline{B}}(\mathbf{r}) &=& 
\frac{1}{8\pi(1-\nu)\mu}\bigg[-(3-4\nu)\ln (r)\, \underline{\hat{\mathbf{I}}}+\frac{\mathbf{r}\mathbf{r}}{r^2}
\nonumber\\&&
\quad{}+(3-4\nu)\ln (R)\, \underline{\hat{\mathbf{I}}}-\frac{\mathbf{R}\,\mathbf{R}}{R^2}\bigg]
\nonumber\\&&{}
+\frac{h}{4\pi(1-\nu)\mu}\frac{1}{R^2}
\bigg[
(\mathbf{\hat{z}}\mathbf{\hat{x}}+\mathbf{\hat{x}}\mathbf{\hat{z}})R_x
\nonumber\\[.1cm]&&\quad{}
-\frac{R_z-h}{3-4\nu}
\bigg\{-
\frac{R_x^2-R_z^2}{R^2}\,\underline{\mathbf{\hat{I}}}
\nonumber\\&&\quad\quad\quad{}
+2(\mathbf{\hat{x}}\mathbf{\hat{z}}-\mathbf{\hat{z}}\mathbf{\hat{x}})\frac{R_xR_z}{R^2}
\bigg\}\bigg]
.
\label{eq_Bexplicit}
\end{eqnarray}

From here, we see that the overall expression decays with the inverse distances at least as $\sim r^{-1}$ and $\sim R^{-1}$ for $r,R\gg h$. This follows for the logarithmic terms from inserting Eq.~(\ref{eq_R}) and combining 
\begin{equation}\label{eq_ln}
\ln(r)-\ln(R) = \ln\frac{r}{R}=\ln\frac{|\mathbf{x}-\mathbf{x}_0|}{|\mathbf{x}-\mathbf{x}_0+2h\mathbf{\hat{z}}|}.
\end{equation} 
By Taylor expansion with respect to $h$, we find that this expression drops with increasing distance from the force center to leading order as $\sim h/r$, that is $\sim r^{-1}$. Therefore, we have formally demonstrated that it is the image counterforce that stabilizes the solution and suppresses the logarithmic divergence. The image provides the necessary counterforce so that for the whole system, real and image contributions together, the net force exerted onto the two-dimensional system vanishes. Physically, it is the boundary that absorbs the net real force and keeps the medium in place.

\section{Effect of the no-slip boundary on the displaceability and mobility of an inclusion}
	\label{sec_displ}
	
We here assume a rigid spherical (actually disk-like) inclusion in the surrounding two-dimensional continuous medium with no-slip surface conditions. \rc{In principle, this means that infinitely large mechanical moduli are assumed for the inclusions. Often, this is a reasonable approximation for hard inclusions in soft surroundings. Considering inclusions of finite elasticity will be an important and interesting, yet much more challenging future task \cite{eshelby1957determination}.} 

In an unbounded system, we can formally calculate the displacement field $\mathbf{u}(\mathbf{x})$ in the elastic environment in response to a force $\mathbf{F}$ acting on the rigid inclusion centered at $\mathbf{x}_0$ via
	\begin{equation}\label{eq_ux}
		\mathbf{u}(\mathbf{x}) = \left(1+\frac{a^2}{4}\nabla^2\right)\underline{\mathbf{G}}(\mathbf{x}-\mathbf{x}_0)\cdot\mathbf{F},
	\end{equation}
where $a$ is the radius of the disk \cite{richter2022mediated}. The expression satisfies Eq.~(\ref{eq_NC})\rc{. Moreover, it is constant on the circumference of the rigid disk, that is, when evaluated for $|\mathbf{x}-\mathbf{x}_0|=a$. This is a necessary requirement to satisfy the no-slip boundary condition on the surface of the inclusion. According to the no-slip condition, the elastic material attached directly to the surface of the disk needs to displace in the same way as the inclusion itself, and, since the inclusion is rigid, the displacement field must be constant along the circumference of the disk. Together, as both Eq.~(\ref{eq_NC}) and the surface condition are satisfied,} Eq.~(\ref{eq_ux}) represents the formal solution. Furthermore, this expression serves to calculate 
the contribution $\mathbf{U}^{(0)}$ to the displacement of the inclusion that results from the term $\mathbf{\underline{G}}(\mathbf{r})$ in Eq.~(\ref{eq_ansatzB}). To this end, Eq.~(\ref{eq_ux}) is evaluated on the surface of the disk, which leads to \cite{richter2022mediated}
\begin{equation}
\mathbf{U}^{(0)} = \frac{1}{8\pi(1-\nu)\mu}\left(
\frac{1}{2}-(3-4\nu)\ln(a)\right)\mathbf{F}. 
\label{eq_U0}
\end{equation}

In the presence of the boundary, the inclusion is additionally exposed to the influence of the image system as given by the terms $-\mathbf{\underline{G}}(\mathbf{R})$ and $\mathbf{\underline{W}}(\mathbf{R})$ in Eq.~(\ref{eq_ansatzB}). 
We evaluate the resulting fields at position $\mathbf{x}_0$ to obtain the corresponding contribution $\mathbf{U}^{(1)}$ to the displacement of the disk. 
To lowest order in $a/h$, we find
\begin{eqnarray}
\mathbf{U}^{(1)} &=&
\frac{1}{8\pi(1-\nu)\mu} \bigg\{
(3-4\nu)\ln(2h)\,\underline{\mathbf{\hat{I}}} - \mathbf{\hat{z}}\mathbf{\hat{z}}
\nonumber\\&&{}\quad\quad
-\frac{1}{2(3-4\nu)}\underline{\mathbf{\hat{I}}}
\bigg\}\cdot\mathbf{F}
+\mathcal{O}\left(\left(\frac{a}{h}\right)^{\!2}\right).
\quad\quad
\label{eq_U1ns}
\end{eqnarray}


Equations (\ref{eq_U0}) and (\ref{eq_U1ns}) both formally contain logarithmic functions of unscaled lengths. However, this ambiguity is resolved when combining the two contributions. We then find for the displacement $\mathbf{U}$ of the disk
\begin{eqnarray}
\mathbf{U}&=&\mathbf{U}^{(0)}+\mathbf{U}^{(1)}
\nonumber\\
&=&
\frac{1}{8\pi(1-\nu)\mu} \bigg\{
(3-4\nu)\ln\left(\frac{2h}{a}\right)\underline{\mathbf{\hat{I}}}
+\frac{1-2\nu}{3-4\nu}\,\underline{\mathbf{\hat{I}}}
\nonumber\\
&&{}
\quad\quad
-\mathbf{\hat{z}}\mathbf{\hat{z}}
\bigg\}\cdot\mathbf{F}
+\mathcal{O}\left(\left(\frac{a}{h}\right)^{\!2}\right).
\label{eq_U}
\end{eqnarray}
Therefore, we observe that the image system and thus the boundary does not only stabilize the induced displacement field, see Eq.~(\ref{eq_ln}) and comments thereafter, but also renders the expression $\mathbf{U}^{(0)}$ for the initial displacement of the disk meaningful. By this statement, we refer to the now dimensionless argument of the logarithm. Formally and illustratively, it is the counterforce contained in the image system and included by the contribution $-\underline{\mathbf{G}}(\mathbf{R})$ in Eq.~(\ref{eq_ansatzB}) that leads to this regularization. 
	
%
For elastic systems, the displaceability matrix $\underline{\mathbf{M}}$ is defined via $\mathbf{U}=\underline{\mathbf{M}}\cdot\mathbf{F}$ \cite{puljiz2016forces, puljiz2017forces} (in analogy to the mobility matrix in low-Reynolds-number hydrodynamics \cite{schmitz1982mobility, brady1988stokesian, dhont1996introduction, guazzelli2012physical, daddi2017mobility, hoell2017dynamical}). In the present context, it is diagonal, that is, $M_{xz}=M_{zx}=0$. From Eq.~(\ref{eq_U}), we read off
\begin{eqnarray}
M_{xx}&=&
\frac{1}{8\pi(1-\nu)\mu} \bigg\{
(3-4\nu)\ln\left(\frac{2h}{a}\right)
+\frac{1-2\nu}{3-4\nu} \bigg\}
\nonumber\\&&\qquad\qquad{}
+\mathcal{O}\left(\left(\frac{a}{h}\right)^{\!2}\right),
\label{eq_Mxx_noslip}
\\
M_{zz}&=&
\frac{1}{8\pi(1-\nu)\mu} \bigg\{
(3-4\nu)\ln\left(\frac{2h}{a}\right)
+2\frac{\nu-1}{3-4\nu} \bigg\}
\nonumber\\&&\qquad\qquad{}
+\mathcal{O}\left(\left(\frac{a}{h}\right)^{\!2}\right).
\label{eq_Mzz_noslip}
\end{eqnarray}

\begin{figure}
 		\centering 
\includegraphics[width=8.5cm]{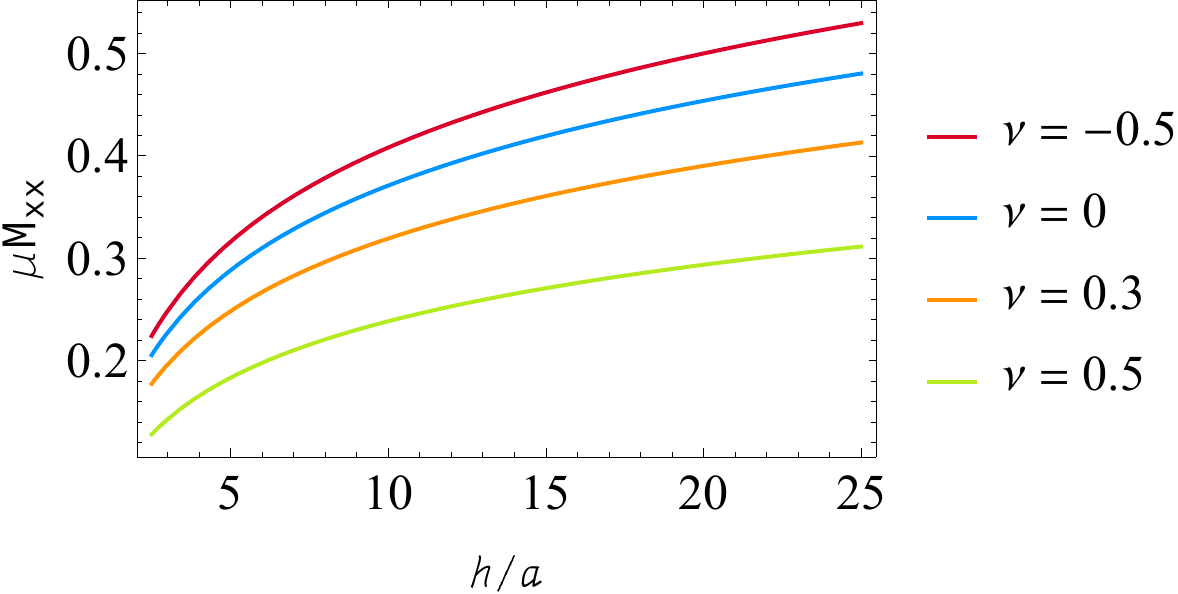}
 		\caption{Displaceability $M_{xx}$ for the displacement of a rigid disk within a two-dimensional elastic membrane of shear modulus $\mu$ and Poisson ratio $\nu$, see Eq.~(\ref{eq_Mxx_noslip}). The disk of radius $a$ is located at a distance $h$ above a flat no-slip boundary. $M_{xx}$ quantifies the displacement parallel to the boundary in response to a force applied to the disk parallel to the boundary.}
 		\label{Fig:displaceability_para}
 	\end{figure}

\begin{figure}
 		\centering 
 		\includegraphics[width=8.5cm]{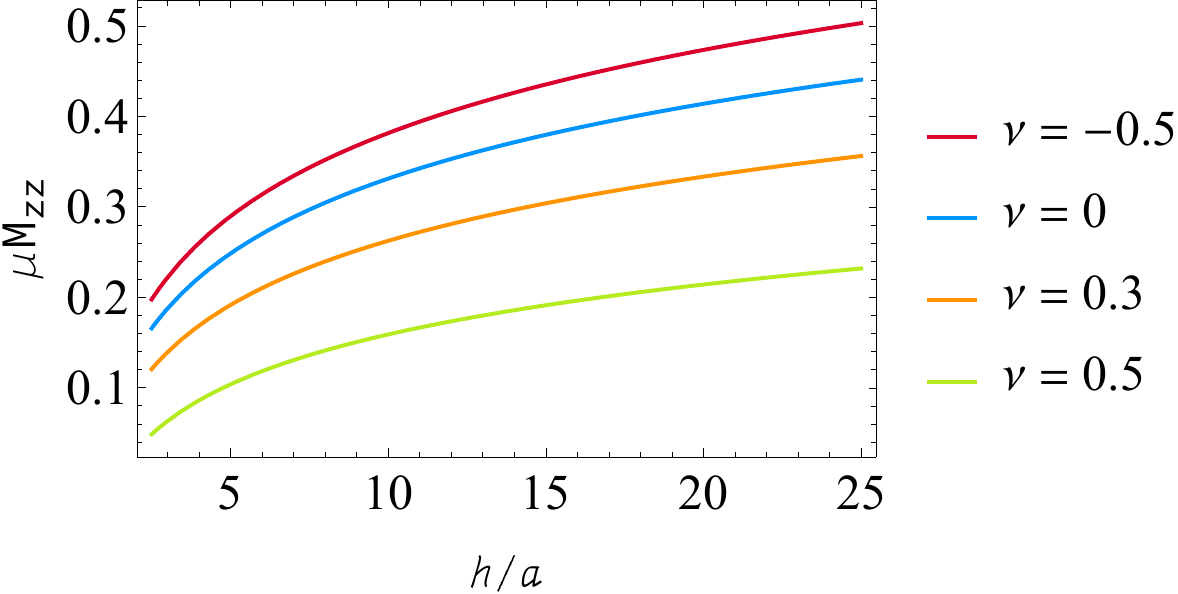}
 		\caption{Same as in Fig.~\ref{Fig:displaceability_para}, yet for the displaceability $M_{zz}$, see Eq.~(\ref{eq_Mzz_noslip}). $M_{zz}$ quantifies the displacement perpendicular to the no-slip boundary in response to a force applied to the disk perpendicular to the boundary.}
 		\label{Fig:displaceability_perp}
 	\end{figure}

First, we note that the displaceability of the inclusion rises as the distance from the no-slip boundary $h/a$ increases. Qualitatively, this behavior is expected, because the pinning impact of the no-slip boundary decreases with increasing distance of the inclusion from this boundary. Second,
we infer that the displaceability $M_{xx}$ associated with forces $\mathbf{F}=F\mathbf{\hat{x}}$ parallel to the no-slip boundary is always larger than $M_{zz}$ linked to forces $\mathbf{F}=F\mathbf{\hat{z}}$ perpendicular to it. 
The displaceabilities 
are plotted in Figs.~\ref{Fig:displaceability_para} and \ref{Fig:displaceability_perp} 
for various values of the Poisson ratio $\nu$.%
%
(To obtain the corresponding mobilities in low-Reynolds-number fluid flows of incompressible liquids, we have to select the value $\nu=1/2$.)

	

\section{Role of the no-slip boundary concerning mediated interactions between inclusions}
\label{sec_mediated}

A similar picture emerges when we consider the displacement $\mathbf{U}^{\mathrm{pair}}$ that a force $\mathbf{F}$ acting on one disk located at position $\mathbf{x}_0$ induces on another inclusion positioned at $\mathbf{x}$. Since the elastic environment is distorted by the action on the first disk, other inclusions confined by the membrane are exposed to these distortions and subject to associated displacements. Thus, they are relocated by the interactions mediated by the elastic surroundings. This effect is quantified by the so-called pair displaceabilities $\underline{\mathbf{M}}^{\mathrm{pair}}$ defined via $\mathbf{U}^{\mathrm{pair}}=\underline{\mathbf{M}}^{\mathrm{pair}}\cdot\mathbf{F}$ \cite{puljiz2016forces, puljiz2017forces, menzel2017force}. (In low-Reynolds-number hydrodynamics, the analogous quantities are the common pair mobilities \cite{dhont1996introduction}.)

We determine the pair displaceabilities by evaluating the displacement field $\mathbf{u}(\mathbf{x})$ at the position $\mathbf{x}$ of the second inclusion as induced by the force $\mathbf{F}$ acting at position $\mathbf{x}_0$ on the first inclusion. From Eqs.~(\ref{eq_ansatzB})--(\ref{eq_Bexplicit}), we find
\begin{eqnarray}
\mathbf{u}(\mathbf{x}) &=&
\frac{1}{8\pi(1-\nu)\mu} \bigg[
{}-(3-4\nu)\ln\left(\frac{r}{R}\right)\underline{\mathbf{\hat{I}}}
\nonumber\\&&{}\quad\quad
+\frac{\mathbf{r}\mathbf{r}}{r^2}-\frac{\mathbf{R}\mathbf{R}}{R^2} \bigg]\cdot\mathbf{F}
\nonumber\\&&{}
+\frac{h}{4\pi(1-\nu)\mu}\frac{1}{R^2}
\bigg[ (\mathbf{\hat{x}}\mathbf{\hat{z}}+\mathbf{\hat{z}}\mathbf{\hat{x}})R_x
\nonumber\\&&{}\quad\quad
-\frac{R_z-h}{3-4\nu}\bigg\{ 
{}-\frac{R_x^2-R_z^2}{R^2}\,\underline{\mathbf{\hat{I}}}
\nonumber\\&&{}\quad\quad\quad\quad
+2(\mathbf{\hat{x}}\mathbf{\hat{z}}-\mathbf{\hat{z}}\mathbf{\hat{x}})\frac{R_xR_z}{R^2}\bigg\} \bigg]\cdot\mathbf{F}
\nonumber\\&&{}
+\mathcal{O}\left(\left(\frac{a}{h}\right)^{\!2}, \left(\frac{a}{r}\right)^{\!2}\right). 
\label{eq_u_pair}
\end{eqnarray}

\begin{figure}
 		\centering 
\includegraphics[width=8.5cm]{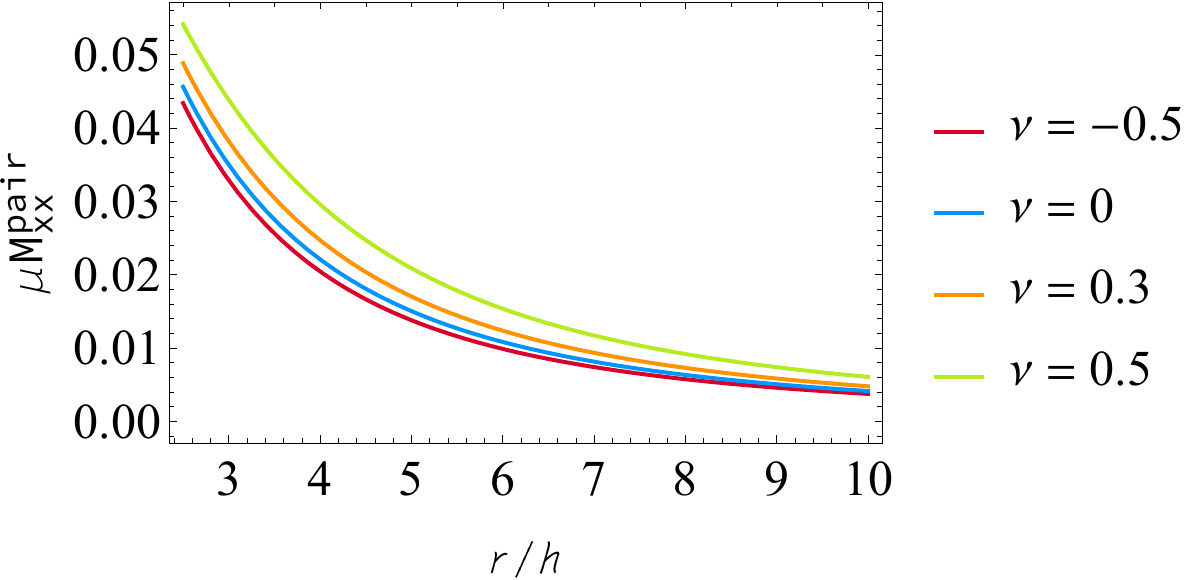}
 		\caption{Pair displaceability $M_{xx}^{\mathrm{pair}}$ quantifying the displacement of one inclusion located at $\mathbf{r}=r\mathbf{\hat{x}}$ from another inclusion within a two-dimensional elastic membrane of shear modulus $\mu$ and Poisson ratio $\nu$, see Eq.~(\ref{eq_Mxx_pair}). Both inclusions are positioned at a distance $h$ above a flat no-slip boundary. $M_{xx}^{\mathrm{pair}}$ quantifies the displacement parallel to the boundary when the force applied to the other inclusion is likewise oriented in parallel direction.}
 		\label{Fig:displaceability_pairxx}
 	\end{figure}

\begin{figure}
 		\centering 
\includegraphics[width=8.5cm]{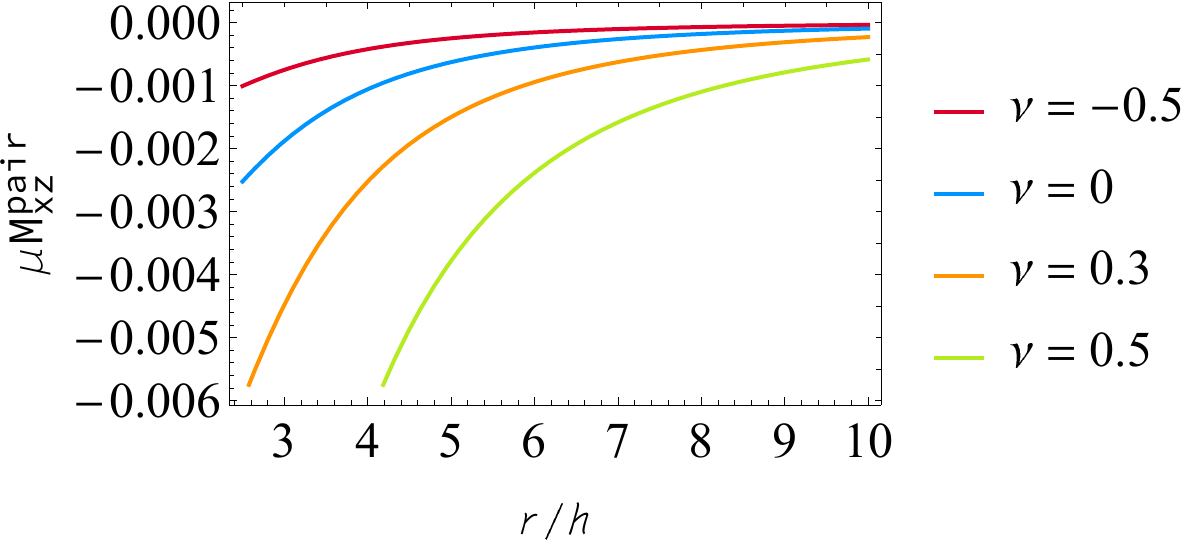}
 		\caption{Same as Fig.~\ref{Fig:displaceability_pairxx}, but for the pair displaceability $M_{xz}^{\mathrm{pair}}$, see Eq.~(\ref{eq_Mxz_pair}). $M_{xz}^{\mathrm{pair}}$ quantifies the displacement parallel to the boundary when the force applied to the other inclusion is oriented in normal direction.}
 		\label{Fig:displaceability_pairxz}
 	\end{figure}

\begin{figure}
 		\centering 
\includegraphics[width=8.5cm]{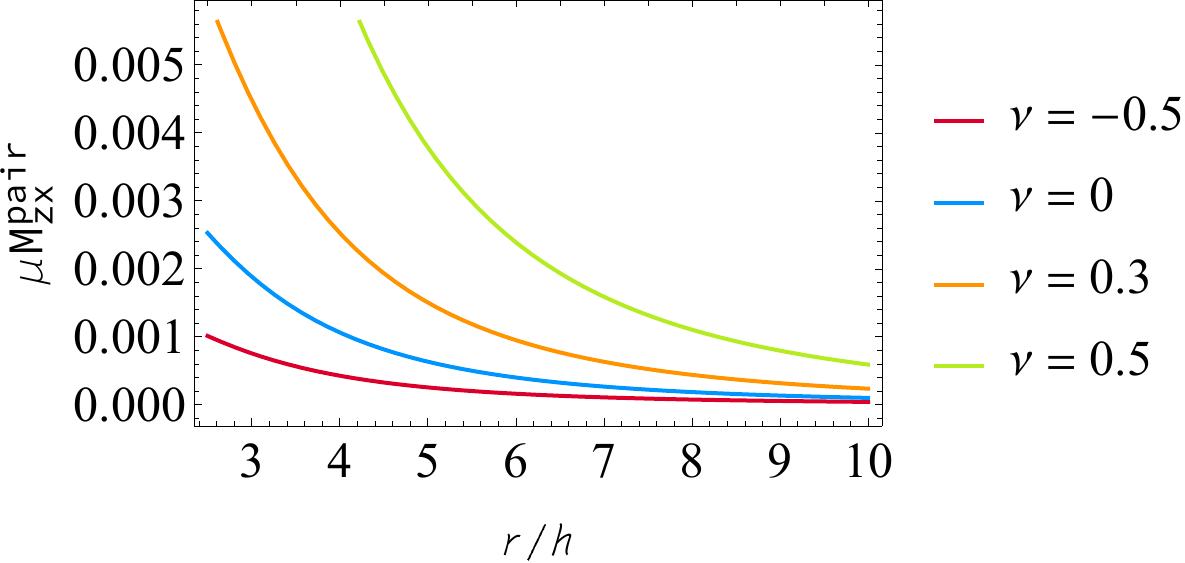}
 		\caption{Same as Fig.~\ref{Fig:displaceability_pairxx}, but for the pair displaceability $M_{zx}^{\mathrm{pair}}$, see Eq.~(\ref{eq_Mzx_pair}). $M_{zx}^{\mathrm{pair}}$ quantifies the displacement normal to the boundary when the force applied to the other inclusion is oriented in parallel direction.}
 		\label{Fig:displaceability_pairzx}
 	\end{figure}

\begin{figure}
 		\centering 
\includegraphics[width=8.5cm]{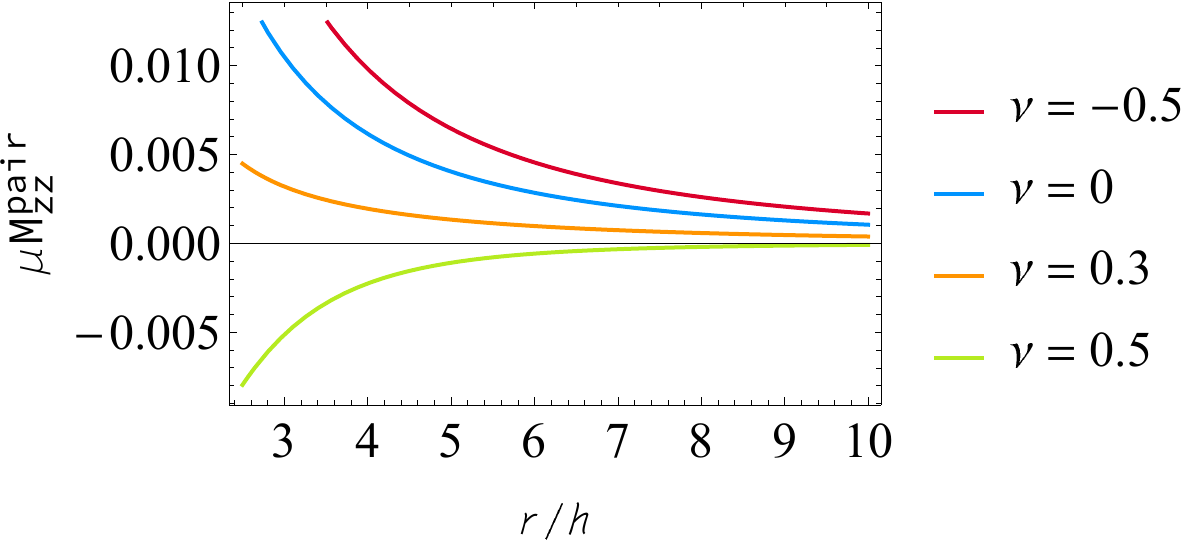}
 		\caption{Same as Fig.~\ref{Fig:displaceability_pairxx}, but for the pair displaceability $M_{zz}^{\mathrm{pair}}$, see Eq.~(\ref{eq_Mzz_pair}). $M_{zz}^{\mathrm{pair}}$ quantifies the displacement normal to the boundary when the force applied to the other inclusion is likewise oriented in normal direction.}
 		\label{Fig:displaceability_pairzz}
 	\end{figure}

In this case, $\mathbf{r}$ denotes the distance vector between the two inclusions, see Eq.~(\ref{eq_rxx0}). We therefore can read off the following pair displaceability matrices from Eq.~(\ref{eq_u_pair}), 
\begin{eqnarray}
M_{xx}^{\mathrm{pair}} &=&
\frac{1}{8\pi(1-\nu)\mu} \bigg[
{}-(3-4\nu)\ln\left(\frac{r}{R}\right)
+\frac{r_x^2}{r^2}
-\frac{R_x^2}{R^2}
\nonumber\\&&{}\quad\quad
+\frac{2h}{3-4\nu}
\frac{R_z-h}{R^2}\frac{R_x^2-R_z^2}{R^2}
\bigg]
\nonumber\\&&{}
+\mathcal{O}\left(\left(\frac{a}{h}\right)^{\!2}, \left(\frac{a}{r}\right)^{\!2}\right), 
\label{eq_Mxx_pair}
\nonumber\\
M_{xz}^{\mathrm{pair}} &=&
\frac{1}{8\pi(1-\nu)\mu} \bigg[
\frac{r_xr_z}{r^2}
-\frac{R_xR_z}{R^2}
\nonumber\\&&{}\quad\quad
+2h\frac{R_x}{R^2}
-\frac{4h(R_z-h)}{3-4\nu}
\frac{R_xR_z}{R^4}
\bigg]
\nonumber\\&&{}
+\mathcal{O}\left(\left(\frac{a}{h}\right)^{\!2}, \left(\frac{a}{r}\right)^{\!2}\right), 
\label{eq_Mxz_pair}
\\
M_{zx}^{\mathrm{pair}} &=&
\frac{1}{8\pi(1-\nu)\mu} \bigg[
\frac{r_xr_z}{r^2}
-\frac{R_xR_z}{R^2}
\nonumber\\&&{}\quad\quad
+2h\frac{R_x}{R^2}
+\frac{4h(R_z-h)}{3-4\nu}
\frac{R_xR_z}{R^4}
\bigg]
\nonumber\\&&{}
+\mathcal{O}\left(\left(\frac{a}{h}\right)^{\!2}, \left(\frac{a}{r}\right)^{\!2}\right), 
\label{eq_Mzx_pair}
\\
M_{zz}^{\mathrm{pair}} &=&
\frac{1}{8\pi(1-\nu)\mu} \bigg[
{}-(3-4\nu)\ln\left(\frac{r}{R}\right)
+\frac{r_z^2}{r^2}
-\frac{R_z^2}{R^2}
\nonumber\\&&{}\quad\quad
+\frac{2h}{3-4\nu}
\frac{R_z-h}{R^2}\frac{R_x^2-R_z^2}{R^2}
\bigg]
\nonumber\\&&{}
+\mathcal{O}\left(\left(\frac{a}{h}\right)^{\!2}, \left(\frac{a}{r}\right)^{\!2}\right). 
\label{eq_Mzz_pair}
\end{eqnarray}
Once more, we infer that the counterforce provided by the image system rescales the argument of the logarithm so that the resulting expressions are well defined. And again we find that to leading order in $h/r$, the magnitudes of the pair interactions decay $\sim r^{-1}$, see also Eq.~(\ref{eq_ln}) and the comments thereafter. It is obvious that the presence of the no-slip boundary breaks the symmetry of the pair displaceability matrix. For two nearby inclusions both located at a distance $h$ above the no-slip boundary, we plot the pair displaceabilities for increasing $r/h$ in Figs.~\ref{Fig:displaceability_pairxx}--\ref{Fig:displaceability_pairzz} for various values of the Poisson ratio. We note the negative values of $M_{zz}^{\mathrm{pair}}$ close to incompressibility, when $\nu$ tends towards $1/2$. There, the upward displacement of one disk away from the boundary requires influx of materials from the sides to approximately maintain the volume, which drags the other disk towards the boundary. 




\section{Free-slip boundary conditions}
\label{sec_freeslip}

A second type of boundary condition that is frequently considered 
is the free-slip case \cite{mathijssen2015tracer, pimponi2016hydrodynamics, menzel2017force, lutz2022tbp}. That is, the fluid or elastic material may not detach from or move into the boundary, but may freely slip along it without any restriction. In our case, we formulate it as
\begin{equation}
u_z(z=0)=0.
\end{equation}
In this situation, the overall Green's function reads
\begin{equation}\label{eq_C}
\mathbf{\underline{C}
}(\mathbf{r})=\mathbf{\underline{G}}(\mathbf{r})+\mathbf{\underline{G}}\!\left(
\mathbf{R}\right)\cdot(\underline{\mathbf{\hat{I}}}-2\mathbf{\hat{z}}\mathbf{\hat{z}}).
\end{equation} 
\rc{For our purposes below, we note that 
the operator $\underline{\mathbf{\hat{I}}}-2\mathbf{\hat{z}}\mathbf{\hat{z}}$ in Eqs.~(\ref{eq_R}) and (\ref{eq_C})
%
%
%
simply implies that the real force and the location of its action are 
mirrored at the free-slip boundary to form a genuine mirror image system.} 
%

As described above, the logarithmic divergence cancels for force configurations in which no net force is exerted on the medium \cite{richter2022mediated}. 
However, 
under free-slip boundary conditions, the no-net-force condition when combining real and image forces only applies if the real force is oriented perpendicular to the boundary\rc{, meaning that $\mathbf{F}=F\mathbf{\hat{z}}$. 
In that case, the first contribution $\mathbf{U}^{(0)}$ to the displacement $\mathbf{U}$ of the inclusion is associated with the first term $\underline{\mathbf{G}}(\mathbf{r})$ on the right-hand side of Eq.~(\ref{eq_C}). It has already been listed in Eq.~(\ref{eq_U0}). The second term on the right-hand side of Eq.~(\ref{eq_C}) results from the image force, and it leads to a contribution $\mathbf{U}^{(1)}$. Since here $\mathbf{r}=\mathbf{x}-\mathbf{x}_0=\mathbf{0}$, we obtain}
\rc{
\begin{eqnarray}
\mathbf{U}^{(1)} &=&
{}-\frac{1}{8\pi(1-\nu)\mu}\bigg[{}1-(3-4\nu) 
\ln(2h)
\bigg]\cdot F\mathbf{\hat{z}}
\nonumber\\[.1cm]&&{}
+\mathcal{O}\bigg(\bigg(\frac{a}{h}\bigg)^{\!\!2}\bigg). 
\label{eq_U1_singlefs}
\end{eqnarray}
}
\rc{Combining Eqs.~(\ref{eq_U0}) and (\ref{eq_U1_singlefs}) leads to the displaceabilities}
\rc{
\begin{eqnarray}
M_{xz} &=& 0,
\label{Mxz_onefs}
\\[.1cm]
M_{zz} &=& \frac{1}{8\pi(1-\nu)\mu}\left[
-\frac{1}{2}-(3-4\nu)\ln\left(\frac{a}{2h}\right)
\right]
\nonumber\\[.1cm]&&
{}+\mathcal{O}\bigg(\bigg(\frac{a}{h}\bigg)^{\!\!2}\bigg).
\label{Mzz_onefs}
\end{eqnarray}
}

Still, the Green's function in Eq.~(\ref{eq_C}) contains the logarithmic dependence $\sim\ln(r)$, if any boundary-parallel component is involved. The mirror image of a real boundary-parallel force component is equal in both magnitude \textit{and orientation} to the real force component, formally resulting in a net boundary-parallel force 
of twice the magnitude instead of a vanishing net force. (Physically, this reflects the ease of slip along the free-slip surface when compared to the situation in which the half-space membrane were anchored at $z=0$.)
 
To stabilize the system, we therefore introduce a second free-slip boundary, here chosen at $x=0$. The system now fills the quarter-space for $x\geq0$ and $z\geq0$. In that case, we obtain the Green's function in the form
\begin{eqnarray}
\mathbf{\underline{C}^{\textrm{quart}}}(\mathbf{r})&=&
\mathbf{\underline{G}}(\mathbf{r}) 
+\mathbf{\underline{G}}\!\left(\mathbf{x}-(\underline{\mathbf{\hat{I}}}-2\mathbf{\hat{x}}\mathbf{\hat{x}})\cdot\mathbf{x}_0\right)\cdot(\underline{\mathbf{\hat{I}}}-2\mathbf{\hat{x}}\mathbf{\hat{x}}) 
\nonumber\\
&&{} +\mathbf{\underline{G}}\!\left(\mathbf{x}-(\underline{\mathbf{\hat{I}}}-2\mathbf{\hat{z}}\mathbf{\hat{z}})\cdot\mathbf{x}_0\right)\cdot(\underline{\mathbf{\hat{I}}}-2\mathbf{\hat{z}}\mathbf{\hat{z}})
\nonumber\\
&&{}+\mathbf{\underline{G}}\!\left(\mathbf{x}-(\underline{\mathbf{\hat{I}}}-2\mathbf{\hat{x}}\mathbf{\hat{x}})\cdot(\underline{\mathbf{\hat{I}}}-2\mathbf{\hat{z}}\mathbf{\hat{z}})\cdot\mathbf{x}_0\right) \nonumber\\
&&\qquad\quad{}\cdot(\underline{\mathbf{\hat{I}}}-2\mathbf{\hat{x}}\mathbf{\hat{x}})\cdot(\underline{\mathbf{\hat{I}}}-2\mathbf{\hat{z}}\mathbf{\hat{z}}).
\label{eq_Cquart}
\end{eqnarray}
%
%
\rc{Here, the leading term $\mathbf{\underline{G}}(\mathbf{r})$ refers to the displacements induced by the real, physical force acting on the material in the quarter-space of $x\geq0$ and $z\geq0$. The next term in $\mathbf{\underline{G}}$ introduces the consequences of an image force generated by mirroring the physical force at the plane $x=0$, to satisfy the free-slip boundary condition at $x=0$. Third, the subsequent contribution in $\mathbf{\underline{G}}$ refers to an image force obtained by mirroring the physical force at the plane $z=0$, to there satisfy the free-slip boundary condition. Both of these two image forces constructed by mirroring the physical force at the planes $x=0$ and $z=0$ are then additionally mirrored at the mutually other plane, to also there satisfy the free-slip boundary condition. Both mirroring steps lead to the same additional image force center with the same additional image force in the quarter-space $x<0$ and $z<0$. Its consequences are included by the last term in $\mathbf{\underline{G}}$ in Eq.~(\ref{eq_Cquart}). In this way, the image system is closed and complete. All boundary conditions are satisfied \cite{lutz2022tbp}.}

This setup generates a force configuration of zero net force, regardless of the orientation of the point force in the real quarter-space. Any pair of real force and its image across the $z=0$ boundary are counterbalanced by the respective images across the $x=0$ boundary. Likewise, any real force and its image across the $x=0$ boundary are counterbalanced by the respective images across the $z=0$ boundary. In other words, any force component parallel to $\mathbf{\hat{x}}$ or $\mathbf{\hat{z}}$ is perpendicular to $\mathbf{\hat{z}}$ or $\mathbf{\hat{x}}$, respectively, which guarantees the overall force-free nature of the combination of all four real and image forces and cancels the logarithmic divergence. 

	\begin{figure}
 		\centering 
\includegraphics[width=8.5cm]{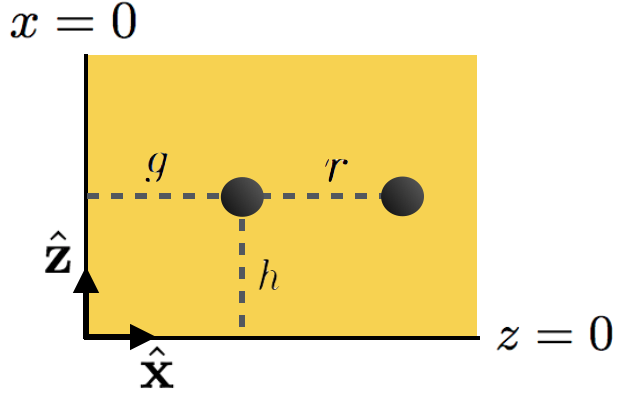}
 		\caption{\rc{Illustration of the geometry considered in the case of two orthogonal free-slip boundaries at $x=0$ and $z=0$. A disk-like inclusion at the position $\mathbf{x}_0=(g,h)$ is embedded in a continuous medium that fills the quarter-space described by $x\geq0$ and $z\geq0$. When illustrating the pair displaceabilities in Fig.~\ref{Fig:Mxxpair_gh1} below, we introduce another inclusion located at a distance vector $\mathbf{r}=r\mathbf{\hat{x}}$ from the first one.}}
 		\label{Fig:schemeII}
 	\end{figure}

Next, we consider a disk-like inclusion located in the membrane at position $\mathbf{x}_0=(g,h)$\rc{, as illustrated in Fig.~\ref{Fig:schemeII}}. Thus, $g$ and $h$ measure the distances relative to the two perpendicular free-slip boundaries.
Moreover, we introduce the abbreviations
\begin{eqnarray}
\mathbf{R}_g&=&\mathbf{r}+2g\mathbf{\hat{x}},
\\
\mathbf{R}_h&=&\mathbf{r}+2h\mathbf{\hat{z}},
\\
\mathbf{R}_{gh}&=&\mathbf{r}+2g\mathbf{\hat{x}}+2h\mathbf{\hat{z}}.
\end{eqnarray}
We determine the displacement $\mathbf{U}$ of the disk when subject to a net force $\mathbf{F}$ in such a geometry. 
The first contribution $\mathbf{U}^{(0)}$ resulting from the term $\underline{\mathbf{G}}(\mathbf{r})$ in Eq.~(\ref{eq_Cquart}) is not related to the presence of the boundaries and therefore takes the same form as in Eq.~(\ref{eq_U0}). Contrarily, the remaining terms in Eq.~(\ref{eq_Cquart}) describe the influence of the image system, 
which we summarize by the contribution $\mathbf{U}^{(1)}$. 
Explicitly, we derive from the remaining terms in Eq.~(\ref{eq_Cquart})
\begin{eqnarray}
\mathbf{U}^{(1)} &=&
\frac{1}{8\pi(1-\nu)\mu}\bigg[{}-(3-4\nu)\bigg\{ 
\ln(2g)\left(\begin{array}{cc}-1&0\\0&1\end{array}\right)
\nonumber\\&&{}\quad\quad
+\ln(2h)\left(\begin{array}{cc}1&0\\0&-1\end{array}\right)
\nonumber\\&&{}\quad\quad
+\ln(2g+2h)\left(\begin{array}{cc}-1&0\\0&-1\end{array}\right)\bigg\}
\nonumber\\[.1cm]&&{}\quad
\rc{-}\,\underline{\mathbf{\hat{I}}}\,
\rc{-}\,\frac{g^2\mathbf{\hat{x}}\mathbf{\hat{x}}
+h^2\mathbf{\hat{z}}\mathbf{\hat{z}}
+gh(\mathbf{\hat{x}}\mathbf{\hat{z}}+\mathbf{\hat{z}}\mathbf{\hat{x}})}
{g^2+h^2}
\bigg]\cdot\mathbf{F}
\nonumber\\[.1cm]&&{}
+\mathcal{O}\bigg(\bigg(\frac{a}{g}\bigg)^{\!\!2},\bigg(\frac{a}{h}\bigg)^{\!\!2}\bigg). 
\label{eq_U1}
\end{eqnarray}

Together, from Eqs.~(\ref{eq_U0}) and (\ref{eq_U1}), as well as from the definition $\mathbf{U}=\underline{\mathbf{M}}\cdot\mathbf{F}$, we find the entries of the displaceability matrix
\begin{eqnarray}
M_{xx} &=&
\frac{1}{8\pi(1-\nu)\mu}
\bigg[ 
\rc{-\frac{1}{2}\,-}\frac{g^2}{g^2+h^2}
\nonumber\\&&{}\quad
-(3-4\nu)\bigg\{
\ln\left(\frac{a}{2g}\right)+\ln\left(\frac{h}{g+h}\right)
\bigg\}
\bigg]
\nonumber\\&&{}
+\mathcal{O}\bigg(\bigg(\frac{a}{g}\bigg)^{\!\!2},\bigg(\frac{a}{h}\bigg)^{\!\!2}\bigg),
\\[.1cm]
M_{xz} &=& M_{zx} = 
{}\rc{{}-}\frac{1}{8\pi(1-\nu)\mu}\frac{gh}{g^2+h^2}
\nonumber\\&&{}\qquad\quad
+\mathcal{O}\bigg(\bigg(\frac{a}{g}\bigg)^{\!\!2},\bigg(\frac{a}{h}\bigg)^{\!\!2}\bigg),
\label{Mxz_fs}
\end{eqnarray}
\begin{eqnarray}
M_{zz} &=&
\frac{1}{8\pi(1-\nu)\mu}
\bigg[ 
\rc{-\frac{1}{2}\,-}\frac{h^2}{g^2+h^2}
\nonumber\\&&{}\quad
-(3-4\nu)\bigg\{
\ln\left(\frac{a}{2h}\right)+\ln\left(\frac{g}{g+h}\right)
\bigg\}
\bigg]
\nonumber\\&&{}
+\mathcal{O}\bigg(\bigg(\frac{a}{g}\bigg)^{\!\!2},\bigg(\frac{a}{h}\bigg)^{\!\!2}\bigg).
\label{Mzz_fs}
\end{eqnarray}
Again, the influence of the image system via $\mathbf{U}^{(1)}$ rescales the argument of the logarithm in $\mathbf{U}^{(0)}$. Moreover, we observe that $M_{xx}$ and $M_{zz}$ follow from each other by switching $g$ and $h$, which is consistent with the underlying geometry. 
\rc{If in Eqs.~(\ref{eq_U1}), (\ref{Mxz_fs}), and (\ref{Mzz_fs}) we let $g/h\rightarrow\infty$, meaning that our force center is located infinitely far away from the free-slip boundary at $x=0$ relative to the boundary at $z=0$, then, for a force $\mathbf{F}=F\mathbf{z}$, we correctly obtain as a limit Eqs.~(\ref{eq_U1_singlefs}), (\ref{Mxz_onefs}), and (\ref{Mzz_onefs}), respectively. In other words, we correctly recover the situation of only one free-slip boundary in the presence of a boundary-normal force.}

\begin{figure}
	\centering 
 	\includegraphics[width=8.5cm]{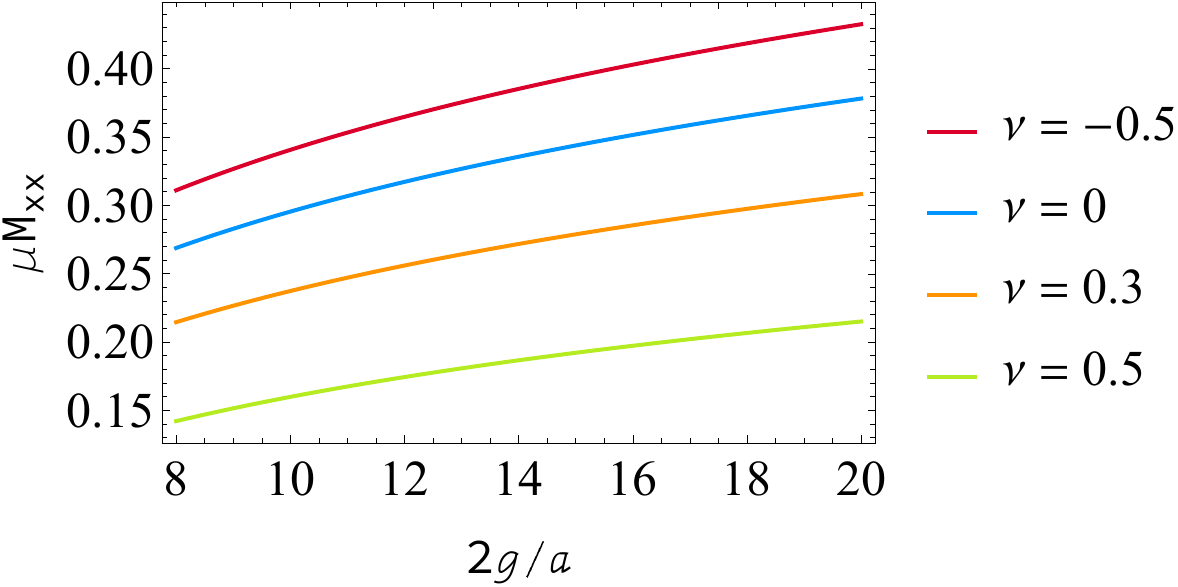}
    \caption{A membrane is confined by two free-slip boundaries located at $x=0$ and $z=0$, while it contains a rigid disk-like inclusion of radius $a$ positioned at $(g,h)$. We depict as a function of $2g/a$ the displaceability $M_{xx}$ of the disk along the $x$-direction for a force likewise applied along the $x$-direction, while we fix $h/g=2$. That is, the disk is positioned closer to the free-slip boundary at $x=0$ than to the free-slip boundary at $z=0$.}
	\label{Fig:fs_Mxx_a-g_1}
\end{figure}
\begin{figure}
	\centering 
 	\includegraphics[width=8.5cm]{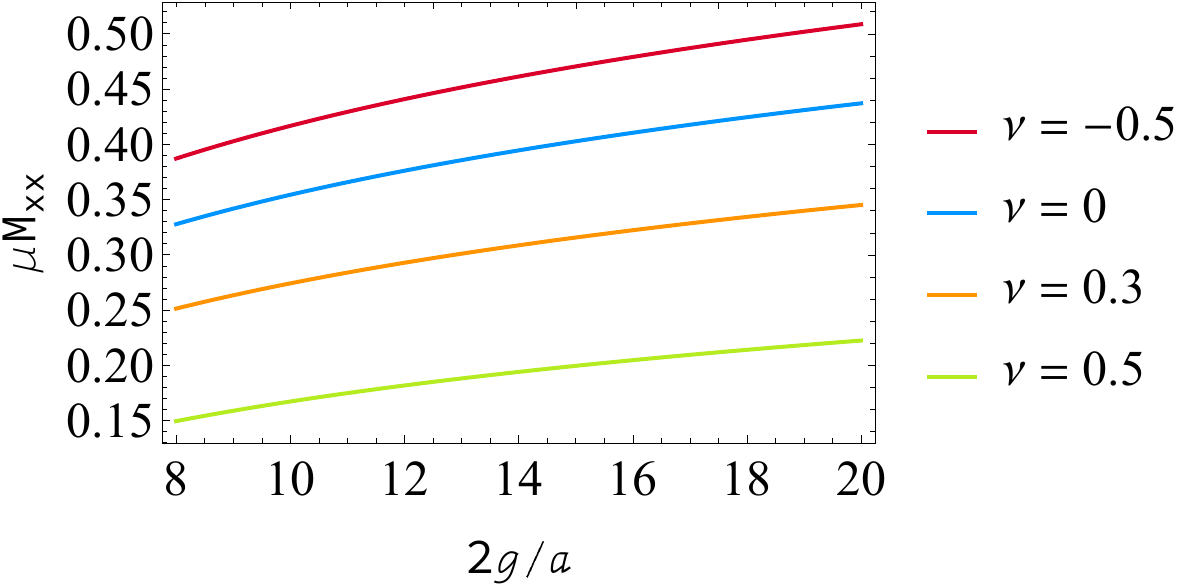}
    \caption{Same as in Fig.~\ref{Fig:fs_Mxx_a-g_1} but for $h/g=0.5$. That is, the disk is positioned closer to the free-slip boundary at $z=0$ than the free-slip boundary at $x=0$.} 
	\label{Fig:fs_Mxx_a-g_2}
\end{figure}
\begin{figure}
	\centering 
 	\includegraphics[width=8.5cm]{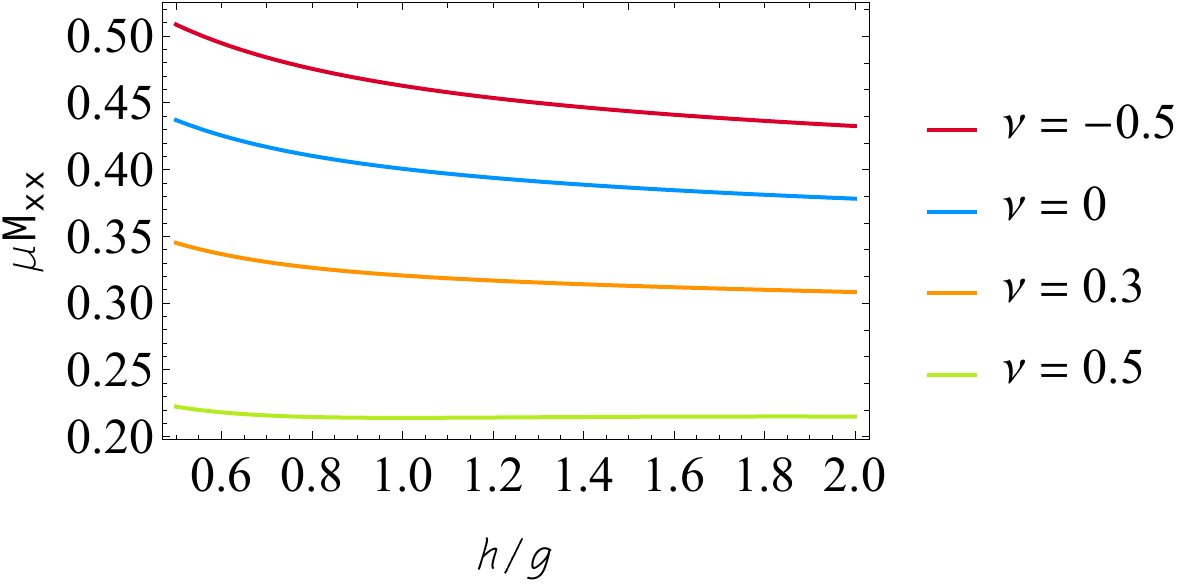}
    \caption{For the same geometry as in Figs.~\ref{Fig:fs_Mxx_a-g_1} and \ref{Fig:fs_Mxx_a-g_2}, we now plot $M_{xx}$ for fixed $2g/a=20$ but varying $h/g$. That is, the vertical distance above the boundary at $z=0$ is increased, while the distance from the boundary at $x=0$ is kept constant.}
	\label{Fig:fs_Mxx_g-h}
\end{figure}

We illustrate in Figs.~\ref{Fig:fs_Mxx_a-g_1}--\ref{Fig:fs_Mxx_g-h} the form of $M_{xx}$ as a function of $a$, $g$, and $h$ for various Poisson ratios $\nu$.
In Figs.~\ref{Fig:fs_Mxx_a-g_1} and \ref{Fig:fs_Mxx_a-g_2}, the distances from both free-slip boundaries increase with increasing $2g/a$, because $h/g$ is kept constant. The considered displacement is normal to the boundary at $x=0$, which confines the displaceability. Thus an increasing distance from $x=0$ supports $M_{xx}$. Simultaneously, an increasing distance from $z=0$ reduces displacements parallel to this free-slip boundary, because more elastic material needs to be taken along. Apparently, from Figs.~\ref{Fig:fs_Mxx_a-g_1} to \ref{Fig:fs_Mxx_a-g_2}, both effects together lead to an increase in $M_{xx}$. Figure~\ref{Fig:fs_Mxx_g-h} exposes the second effect, where we keep $g$ constant and only increase the vertical distance $h$ from the free-slip boundary at $z=0$. Then, the mobility $M_{xx}$ decreases as described above. 
%
%

For $M_{xz}=M_{zx}$, the Poisson ratio only affects the magnitude of the displaceability. Since, in general, $M_{xz}=M_{zx}\neq0$, the two orthogonal free-slip boundaries induce displacements normal to boundary-parallel force components. The maximum of this effect occurs for $g=h$. Finally, the behavior of $M_{zz}$ is identical to the one of $M_{xx}$ for switched $g$ and $h$.

\section{Mediated interactions between inclusions near perpendicular free-slip boundaries}\label{sec_free-slip_mediated}

In analogy to Sec.~\ref{sec_mediated}, we calculate the displacement $\mathbf{U}^{\mathrm{pair}}$ of an inclusion at position $\mathbf{x}$ in response to the force $\mathbf{F}$ acting on another disk located at position $\mathbf{x}_0$. Again, the pair displaceabilites $\underline{\mathbf{M}}^{\mathrm{pair}}$ are defined via $\mathbf{U}^{\mathrm{pair}} = \underline{\mathbf{M}}^{\mathrm{pair}}\cdot\mathbf{F}$. 
\rc{For reference, we begin as in Sec.~\ref{sec_freeslip} by considering just one free-slip boundary located at $z=0$, so that the Green's function $\underline{\mathbf{C}}(\mathbf{r})$ in Eq.~(\ref{eq_C}) applies, together with a boundary-normal force $\mathbf{F}=F\mathbf{\hat{z}}$. $\mathbf{r}$ now marks the distance vector from the first inclusion, onto which the force $\mathbf{F}$ is acting, to the second inclusion, which is exposed to the resulting displacements in the surrounding medium.}

\rc{
As in Sec.~\ref{sec_mediated}, we determine the pair displaceabilities from the displacement field $\mathbf{u}(\mathbf{x})$ at position $\mathbf{x}$ as induced by the force $\mathbf{F}$ acting on the disk at position $\mathbf{x}_0$. 
Specifically, from $\mathbf{u}(\mathbf{x})=\mathbf{\underline{C}}(\mathbf{r})\cdot\mathbf{F}$ and Eq.~(\ref{eq_C}), we find 
\begin{eqnarray}
\mathbf{u}(\mathbf{x}) &=&
\frac{1}{8\pi(1-\nu)\mu}\bigg[
{}-(3-4\nu)
\ln\left(\frac{r}{R} \right)
\nonumber\\[.1cm]
&&{}\qquad
+\frac{\mathbf{r}\mathbf{r}}{r^2}
-\frac{\mathbf{R}\mathbf{R}}{R^2}
\bigg]
\!\cdot\! F\mathbf{\hat{z}}
\nonumber\\[.1cm]
&&{}+\mathcal{O}\bigg(\bigg(\frac{a}{h}\bigg)^{\!\!2},\bigg(\frac{a}{r}\bigg)^{\!\!2}\bigg).
\label{eq_u_C}
\end{eqnarray}
}
\rc{From this equation, we directly read off the displaceabilities
\begin{eqnarray}
M_{xz}^{\mathrm{pair}} &=&
\frac{1}{8\pi(1-\nu)\mu} \left[
\frac{r_xr_z}{r^2}-\frac{R_xR_z}{R^2}
\right]
\nonumber\\[.1cm]
&&{}+\mathcal{O}\bigg(\bigg(\frac{a}{h}\bigg)^{\!\!2},\bigg(\frac{a}{r}\bigg)^{\!\!2}\bigg), 
\label{eq_Mxz1fs}
\\[.1cm]
M_{zz}^{\mathrm{pair}} &=&
\frac{1}{8\pi(1-\nu)\mu} \bigg[
{}-(3-4\nu)\ln\frac{r}{R}
\nonumber\\[.1cm]
&&{}\qquad+
\frac{r_zr_z}{r^2}-\frac{R_zR_z}{R^2}
\bigg]
\nonumber\\[.1cm]
&&{}+\mathcal{O}\bigg(\bigg(\frac{a}{h}\bigg)^{\!\!2},\bigg(\frac{a}{r}\bigg)^{\!\!2}\bigg),
\label{eq_Mzz1fs}
\end{eqnarray}
where $\mathbf{R}$ was defined in Eq.~(\ref{eq_R}).}

\rc{Next, we turn to the two free-slip boundaries located at $x=0$ and $z=0$. Thus, we proceed in the analogous way, but now determine the displacement field $\mathbf{u}(\mathbf{x})$ induced by a force $\mathbf{F}$ of arbitrary orientation within the two-dimensional plane. We calculate $\mathbf{u}(\mathbf{x})=\mathbf{\underline{C}^{\textrm{quart}}}(\mathbf{r})\cdot\mathbf{F}$ from Eq.~(\ref{eq_Cquart}) and obtain} 
\begin{eqnarray}
\mathbf{u}(\mathbf{x}) &=&
\frac{1}{8\pi(1-\nu)\mu}\bigg[
{}-(3-4\nu)\bigg\{
\ln(r)\left(\begin{array}{cc}1&0\\0&1\end{array}\right)
\nonumber\\[.1cm]
&&{}\quad\quad+\ln(R_g)\left(\begin{array}{cc}-1&0\\0&1\end{array}\right)
+\ln(R_h)\left(\begin{array}{cc}1&0\\0&-1\end{array}\right)
\nonumber\\[.1cm]
&&{}\quad\quad+\ln(R_{gh})\left(\begin{array}{cc}-1&0\\0&-1\end{array}\right)
\bigg\}
\nonumber\\[.1cm]
&&{}\quad
+\frac{\mathbf{r}\mathbf{r}}{r^2} \rc{ \:\cdot\left(\begin{array}{cc}1&0\\0&1\end{array}\right)}+\frac{\mathbf{R}_g\mathbf{R}_g}{R_g^2}\rc{\:\cdot\left(\begin{array}{cc}-1&0\\0&1\end{array}\right)}\nonumber\\[.1cm]
&&{}\quad
+\frac{\mathbf{R}_h\mathbf{R}_h}{R_h^2}\rc{\:\cdot\left(\begin{array}{cc}1&0\\0&-1\end{array}\right)}\nonumber\\[.1cm]
&&{}\quad+\frac{\mathbf{R}_{gh}\mathbf{R}_{gh}}{R_{gh}^2}\rc{\:\cdot\left(\begin{array}{cc}-1&0\\0&-1\end{array}\right)}\bigg]\cdot\mathbf{F}
\nonumber\\[.1cm]&&
{}+\mathcal{O}\bigg(\bigg(\frac{a}{g}\bigg)^{\!\!2},\bigg(\frac{a}{h}\bigg)^{\!\!2},\bigg(\frac{a}{r}\bigg)^{\!\!2}\bigg).
\label{eq_u_Cquart}
\end{eqnarray}

Again, $\mathbf{r}$ here denotes the distance vector between the disk exposed to $\mathbf{F}$ and the inclusion subject to the resulting displacement in the elastic medium. Thus we directly obtain from Eq.~(\ref{eq_u_Cquart})
\begin{eqnarray}
M_{xx}^{\mathrm{pair}} &=&
\frac{1}{8\pi(1-\nu)\mu}\bigg[
{}-(3-4\nu)\bigg\{
\ln\left(\frac{r}{R_g}\right)
\nonumber\\&&{}\quad\quad
+\ln\left(\frac{R_h}{R_{gh}}\right)\bigg\}
+\frac{r_xr_x}{r^2}
\nonumber\\&&{}\quad\quad
\rc{-}\frac{R_{g,x}R_{g,x}}{R_g^2}
+\frac{R_{h,x}R_{h,x}}{R_h^2}
\rc{-}\frac{R_{gh,x}R_{gh,x}}{R_{gh}^2}\bigg]
\nonumber\\&&{}
+\mathcal{O}\bigg(\bigg(\frac{a}{g}\bigg)^{\!\!2},\bigg(\frac{a}{h}\bigg)^{\!\!2}, \bigg(\frac{a}{r}\bigg)^{\!\!2}\bigg), 
\label{eq_free-slip_Mxxpair}
\\[.1cm]
M_{xz}^{\mathrm{pair}} &=&
\frac{1}{8\pi(1-\nu)\mu}\bigg[
\frac{r_xr_z}{r^2}
+\frac{R_{g,x}R_{g,z}}{R_g^2}
\nonumber\\&&{}\quad\quad\quad\quad
\quad\quad
\rc{-}\frac{R_{h,x}R_{h,z}}{R_h^2}
\rc{-}\frac{R_{gh,x}R_{gh,z}}{R_{gh}^2}\bigg]
\nonumber\\&&{}
+\mathcal{O}\bigg(\bigg(\frac{a}{g}\bigg)^{\!\!2},\bigg(\frac{a}{h}\bigg)^{\!\!2}, \bigg(\frac{a}{r}\bigg)^{\!\!2}\bigg), 
\label{eq_free-slip_Mxzpair}
\\[.1cm]
M_{zx}^{\mathrm{pair}} &=&
\frac{1}{8\pi(1-\nu)\mu}\bigg[
\frac{r_zr_x}{r^2}
\rc{-}\frac{R_{g,z}R_{g,x}}{R_g^2}
\nonumber\\&&{}\quad\quad\quad\quad
\quad\quad
+\frac{R_{h,z}R_{h,x}}{R_h^2}
\rc{-}\frac{R_{gh,z}R_{gh,x}}{R_{gh}^2}\bigg]
\nonumber\\&&{}
+\mathcal{O}\bigg(\bigg(\frac{a}{g}\bigg)^{\!\!2},\bigg(\frac{a}{h}\bigg)^{\!\!2}, \bigg(\frac{a}{r}\bigg)^{\!\!2}\bigg), 
\label{eq_free-slip_Mzxpair}
\\[.1cm]
M_{zz}^{\mathrm{pair}} &=&
\frac{1}{8\pi(1-\nu)\mu}\bigg[
{}-(3-4\nu)\bigg\{
\ln\left(\frac{r}{R_h}\right)
\nonumber\\&&{}\quad\quad
+\ln\left(\frac{R_g}{R_{gh}}\right)\bigg\}
+\frac{r_zr_z}{r^2}
\nonumber\\&&{}\quad\quad
+\frac{R_{g,z}R_{g,z}}{R_g^2}
\rc{-}\frac{R_{h,z}R_{h,z}}{R_h^2}
\rc{-}\frac{R_{gh,z}R_{gh,z}}{R_{gh}^2}\bigg]
\nonumber\\&&{}
+\mathcal{O}\bigg(\bigg(\frac{a}{g}\bigg)^{\!\!2},\bigg(\frac{a}{h}\bigg)^{\!\!2}, \bigg(\frac{a}{r}\bigg)^{\!\!2}\bigg). 
\label{eq_free-slip_Mzzpair}
\end{eqnarray}


One final time, we remark that the argument of the logarithm in Eq.~(\ref{eq_G}) gets rescaled through the presence of the counterforces of the image system. The resulting expressions are well defined. Moreover, to leading order the magnitudes of the pair interactions decay as $\sim g/r$ and $\sim h/r$, see also Eq.~(\ref{eq_ln}) and the comments thereafter. 
\rc{For $\mathbf{F}=F\mathbf{\hat{z}}$, letting $g/h\rightarrow\infty$, Eqs.~(\ref{eq_u_Cquart}), (\ref{eq_free-slip_Mxzpair}), and (\ref{eq_free-slip_Mzzpair}) turn into Eqs.~(\ref{eq_u_C}), (\ref{eq_Mxz1fs}), and (\ref{eq_Mzz1fs}), respectively, for just one free-slip boundary at $z=0$ and $\mathbf{R}_h\equiv\mathbf{R}$, as expected.}

\begin{figure}
	\centering 
\includegraphics[width=8.5cm]{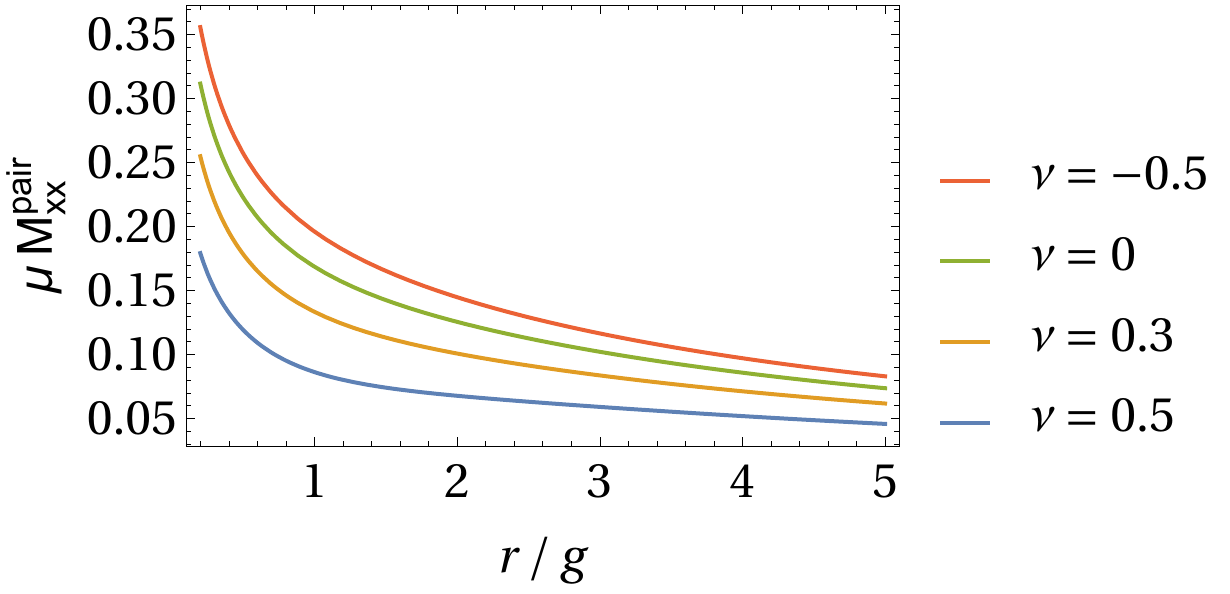}
    \caption{Pair displaceability $M_{xx}^{\mathrm{pair}}$ describing displacements of an inclusion located at $\mathbf{r}=r\mathbf{\hat{x}}$ from another inclusion positioned at $\mathbf{x}_0=(g,h)$ within a two-dimensional elastic membrane of shear modulus $\mu$ and Poisson ratio $\nu$, see Eq.~(\ref{eq_free-slip_Mxxpair}). The membrane is confined by two free-slip boundaries at $x=0$ and at $z=0$.  Considered displacements along the $x$-direction here are induced by a force acting on the inclusion at $\mathbf{x}_0=(g,h)$ likewise along the $x$-direction. In the depicted case, we set $g/h=1$, that is, $\mathbf{x}_0$ is located on the diagonal between the two orthogonal free-slip boundaries.} 
	\label{Fig:Mxxpair_gh1}
\end{figure}

For illustration, we again consider two nearby inclusions, both located at a distance $h$ above the free-slip boundary at $z=0$. Thus, the first inclusion is centered at $\mathbf{x}_0=(g,h)$, while the second one is found at $\mathbf{x}=(g+r,h)$\rc{, see Fig.~\ref{Fig:schemeII}}. We plot the pair displaceability $M_{xx}^{\mathrm{pair}}$ for increasing $r/g$ for $g/h=1$ in Fig.~\ref{Fig:Mxxpair_gh1}. 
%
%
%
Concerning $M_{xz}^{\mathrm{pair}}=M_{zx}^{\mathrm{pair}}$, the Poisson ratio in the expression of Eq.~(\ref{eq_free-slip_Mxzpair}) only scales the magnitude. In the expression for $M_{zz}^{\mathrm{pair}}$ in Eq.~(\ref{eq_free-slip_Mzzpair}), the roles of $g$ and $h$ are again switched when compared to $M_{xx}^{\mathrm{pair}}$.

\section{Conclusions}\label{sec_concl}

In summary, we have addressed the issue of apparent logarithmic divergence of flow and displacement fields in two-dimensional fluid or elastic systems, when they are exposed to concentrated net forces. Example situations are given by small inclusions on which external forces are exerted. Due to the confinement by the continuous environment, these forces are transferred to the surrounding medium. We keep to strictly linear media, that is linearly elastic systems or incompressible liquids under low-Reynolds-number hydrodynamics. 

In fact, the formal logarithmic divergence of the associated Green's function at the end does not constitute a weakness of the linear description. It rather expresses that even an infinitely extended system in two dimensions cannot compensate for a net force. Instead, the net force is transmitted to and compensated by the boundaries of a necessary confinement. 

As we demonstrate, already one no-slip boundary, for example, when clamping a thin elastic membrane, is sufficient for stabilization. The net force transmitted to the boundary is compensated by counterforces emerging from the boundary. In the method of images, it is represented by a suitable image system.
Besides, in the case of a free-slip boundary, a kink must be introduced for stabilization. 

Consequently, the mobilities or displaceabilities of inclusions in the medium are affected. Details depend on the nature of the boundaries, the distance(s) from the boundaries, and whether forces are applied alongside or normal to the boundaries. Likewise, the mutual interactions mediated between inclusions through the fluid or elastic surroundings are affected by the boundaries. 

On the one hand, our results are conceptually important concerning the theoretical description of (flat) thin fluid films and elastic membranes when subject to net forces. They indicate that no unnatural behavior emerges that would point to a problem with the underlying linear equations \cite{proudman1957expansions}. Instead, the role of the boundaries cannot be neglected any longer for such geometries. An interesting question is in which way a flexible, nonlinearly elastic no-slip boundary is likewise capable of stabilizing the logarithmic divergence \cite{daddi2019frequency}. \rc{Another aspect concerns further types of boundary condition, particularly of the partial-slip type \cite{lauga2005brownian}. Although mathematically more involved, the key concept that the boundary needs to stabilize the system in the presence of a net force acting on the material should remain, at least as long as a physically stable situation is addressed and the dynamics remains overdamped.} 

On the other hand, our results will support the quantification of the behavior of inclusions in real set-ups, for instance, in flat anchored or clamped biological or artificial membranes \cite{malmstadt2006automated, urayama2007stretching, han2007nanopore, brommel2013orientation, tarun2018interaction, amador2021hydrodynamic, tokumoto2021probing}. 
\rc{Particularly, we hope to stimulate by our work quantitative experimental investigations and confirmation of our results. To this end, thin free-standing fluid films or elastic membranes of little compressibility along their surface normal, but fluid-like behavior or low to moderate magnitude of elastic moduli in the in-plane directions represent most promising candidates. Free-standing smectic-A liquid-crystalline films \cite{eremin2011two} or membranes of smectic-A liquid-crystalline elastomers \cite{gebhard1998freestanding} of homeotropic director alignment may be used for this purpose. Our results should further be helpful to quantify and optimize the performance of new types of composite membrane materials. For instance, magnetic inclusions in elastic membranes respond directly to additional magnetic field gradients. Thus, the membranes do not require excitation by additional mechanically coupled actuators \cite{raikher2008shape} and could be used as loudspeakers by themselves. Yet, the induced magnetic forces between the inclusions will affect the overall mechanical properties and deformation of the materials as well.}

\begin{acknowledgments}
The authors thank the German Research Foundation (Deutsche Forschungsgemeinschaft, DFG) for support of this work through the grant no.\ ME 3571/5-1. Moreover, A.M.M.\ acknowledges support by the DFG through the Heisenberg grant no.\ ME 3571/4-1. 
\end{acknowledgments}


%

\end{document}